\documentclass{optica-article}

\journal{opticajournal} 

\articletype{Research Article}

\usepackage{lineno}
\usepackage{gensymb}
\usepackage{tabularx}

\begin{document}

\title{Highly efficient visible and near-IR photon pair generation with thin-film lithium niobate}

\author{
Nathan A. Harper,\authormark{1,$\dag$}
Emily Y. Hwang,\authormark{2,$\dag$}
Ryoto Sekine,\authormark{3}
Luis Ledezma,\authormark{3}
Christian Perez,\authormark{4}
Alireza Marandi,\authormark{2,3}
and Scott K. Cushing\authormark{1,*}
}

\address{
\authormark{1}Division of Chemistry and Chemical Engineering, California Institute of Technology, Pasadena, California 91125, USA\\
\authormark{2}Department of Applied Physics and Materials Science, California Institute of Technology, Pasadena, California, 91125, USA\\
\authormark{3}Department of Electrical Engineering, California Institute of Technology, Pasadena, California 91125, USA\\
\authormark{4}California State University, Los Angeles, 90032, USA\\
\authormark{$\dag$}These authors contributed equally to this work.
}

\email{\authormark{*}scushing@caltech.edu} 


\begin{abstract} 
Efficient on-chip entangled photon pair generation at telecom wavelengths is an integral aspect of emerging quantum optical technologies, particularly for quantum communication and computing. However, moving to shorter wavelengths enables the use of more accessible silicon detector technology and opens up applications in imaging and spectroscopy. Here, we present high brightness ($(1.6 \pm 0.3) \times 10^{9}$~pairs/mW/nm) visible-near-IR photon pair generation in a periodically poled lithium niobate nanophotonic waveguide. The degenerate spectrum of the photon pairs is centered at 811~nm with a bandwidth of 117~nm. The measured on-chip source efficiency of $(2.3\pm 0.5) \times 10^{11}$~pairs/mW is on par with source efficiencies at telecom wavelengths and is also orders of magnitude higher than the efficiencies of other visible sources implemented in bulk crystal or diffused waveguide-based technologies. These results represent the shortest wavelength of photon pairs generated in a nanophotonic waveguide reported to date by nearly an octave.
\end{abstract}

\section{Introduction}

Spontaneous parametric downconversion (SPDC) has been used for decades to produce quantum entanglement in various photonic degrees of freedom, serving as a workhorse in emerging quantum optical technologies. Compared to most nonlinear processes, SPDC is relatively inefficient, requiring over one million pump photons to produce one pair of entangled photons in even the highest performing crystals. However, recent advances in nanophotonics, particularly in thin-film lithium niobate (TFLN), have enabled significantly more efficient frequency conversion and quantum state generation\cite{jankowski_ultrabroadband_2020,wang_ultrahigh-efficiency_2018,nehra_few-cycle_2022} through sub-$\mu$m interaction areas, high nonlinearities, and low material losses\cite{boes_lithium_2023,zhu_integrated_2021}. By exploiting this platform, many recent demonstrations of SPDC in TFLN \cite{zhao_high_2020,javid_ultrabroadband_2021,xue_ultrabright_2021,zhang_scalable_2023} have achieved efficiencies three orders of magnitude greater than that of bulk crystal-based sources\cite{dayan_nonlinear_2005} and one order of magnitude greater than that of large diffused waveguide-based sources\cite{bock_highly_2016}. To date, most TFLN-based photon pair sources are designed for SPDC at telecom wavelengths because of the low losses in optical fibers at 1550~nm\cite{miya_ultimate_1979,nagayama_ultra-low-loss_2002} and back-compatibility for applications such as quantum communication\cite{gisin_quantum_2007}, computing\cite{obrien_optical_2007}, and a globally connected quantum network\cite{kimble_quantum_2008}.

Although telecom photons are preferred for quantum information applications, visible and near-infrared photons are generally better suited for imaging and spectroscopy. Experiments at these wavelengths can take advantage of multi-pixel detectors such as electron-multiplying charge coupled devices (EM-CCD) and single photon avalanche detector (SPAD) arrays, enabling the measurements needed for imaging\cite{toninelli_resolution-enhanced_2019,defienne_pixel_2022,moreau_imaging_2019} and characterization of high-dimensional entangled states\cite{zia_interferometric_2023,zhang_high_2021}. Furthermore, the electronic transitions of molecules and atoms are accessible at near-IR wavelengths, allowing for fluorescence lifetime measurements\cite{harper_entangled_2023,eshun_fluorescence_2023,scarcelli_entangled-photon_2008}, compatibility with quantum memories\cite{zhang_preparation_2011}, and fundamental studies of few-photon nonlinearities\cite{tabakaev_spatial_2022,dayan_nonlinear_2005,peer_temporal_2005}. More generally, near-IR and visible photons can be detected with high quantum efficiency and low dark noise using existing mature silicon technology at room temperature, compared to near-IR-IR detectors which require cryogenic cooling\cite{ceccarelli_recent_2021}. Despite these advantages, all demonstrations of nanophotonic pair production have resided in the telecom region, and the best near-IR and visible photon pair sources are still large-area waveguides\cite{fiorentino_spontaneous_2007,tanzilli_highly_2001,gabler_photon_2023,jechow_high_2008} and bulk periodically poled crystals\cite{fiorentino_spontaneous_2007,nasr_ultrabroadband_2008,tanaka_noncollinear_2012,okano_054_2015,dayan_nonlinear_2005,sensarn_generation_2010,tabakaev_spatial_2022}. Potential reasons for this discrepancy stem from the difficulty in fabricating visible nonlinear circuits on thin-film lithium niobate due to factors such as the ultra-short poling periods required for quasi-phase matching and losses from material absorption\cite{bhatt_urbach_2012,schwesyg_light_2010,ciampolillo_quantification_2011} and scattering\cite{maurer_glass_1973}. In spite of these difficulties, high-performance visible devices in thin-film lithium niobate are becoming increasingly common for classical applications such as electro-optic modulation and second harmonic generation\cite{xue_full_spectrum_2023, renaud_sub-1_2023, park_high-efficiency_2022, desiatov_ultrea-low-loss_2019, sayem_efficient_2021}.

\begin{figure}[b]
\centering\includegraphics[width=7cm]{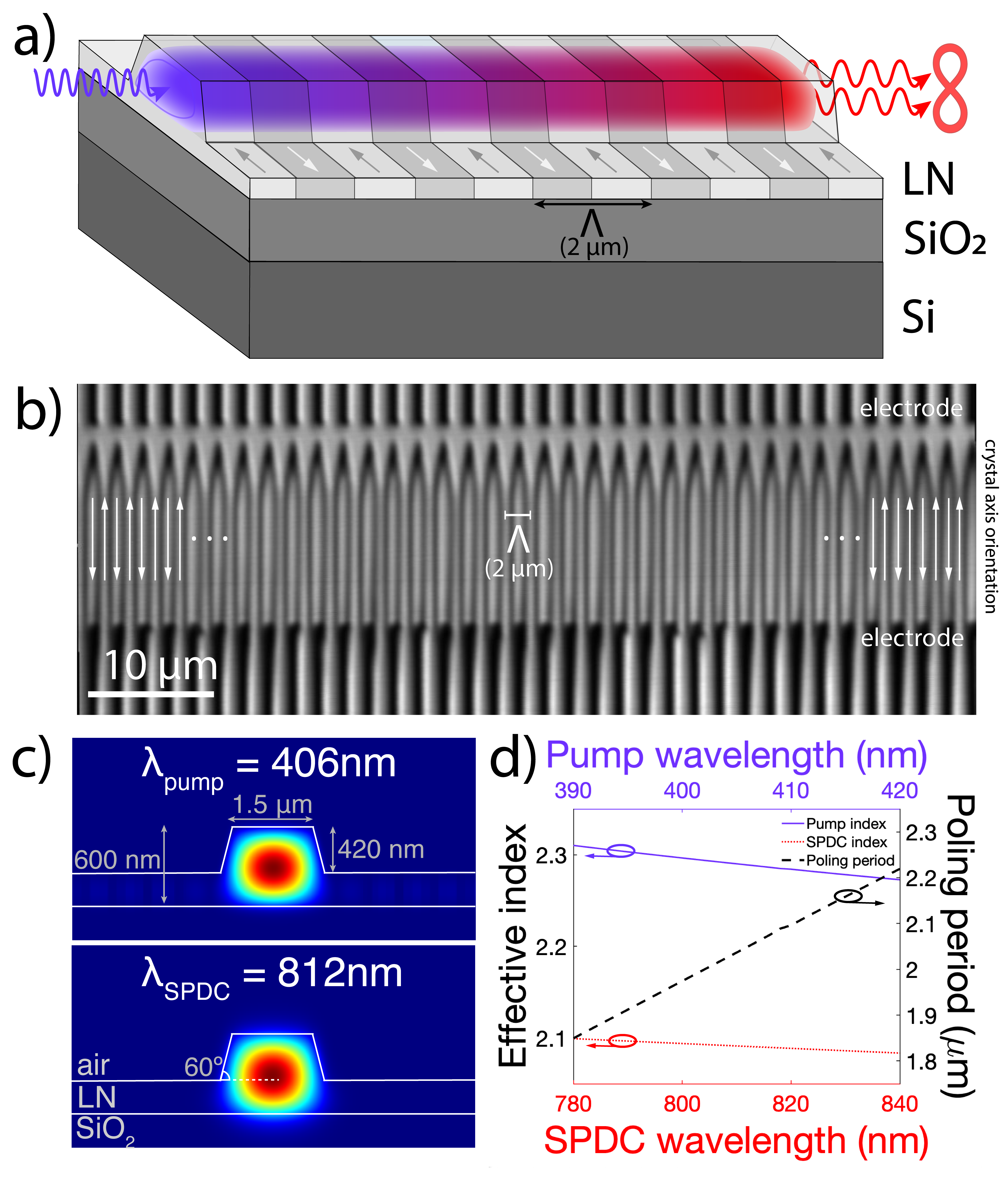}
\caption{\label{fig:design} (a)~Schematic of the periodically poled lithium niobate nanophotonic waveguide. (b)~Second harmonic microscopy image of the periodic poling. The poling electrode period in this image is 2~$\mu$m. (c)~Mode profiles and waveguide geometry of the fundamental quasi-TE modes at the designed pump and SPDC center wavelengths. (d)~Refractive indices and corresponding poling periods for a range of SPDC wavelengths.}
\end{figure}

Here, we extend TFLN-based SPDC sources by nearly an octave in frequency to produce high-brightness photon pairs in the visible and near-IR. The device produces an on-chip efficiency of $(2.3\pm 0.5) \times 10^{11}$ pairs/mW, which corresponds to a per-photon conversion efficiency of more than 1 photon pair converted in every 10,000 pump photons ($(1.1 \pm 0.2) \times 10^{-4} $~pairs/photon). This efficiency is nearly two orders of magnitude better than visible-light diffused waveguide SPDC sources\cite{gabler_photon_2023} and is on par with other TFLN sources in the telecom regime\cite{xue_ultrabright_2021}. The SPDC from this device exhibits a broad spectrum centered at 811~nm with a degenerate FWHM bandwidth of 117~nm and an average brightness of $(1.6 \pm 0.3) \times 10^{9}$~pairs/mW/nm, a number limited by at least an order of magnitude by the pump laser linewidth (0.8~nm). Consistent with this bandwidth, we measure an ultrashort coherence time of $\sim$40~fs for the entangled photons with an indistinguishability of $100. \pm 1\%$. Our results therefore show that, although pumped at wavelengths near what would usually be considered its cutoff range, TFLN can equally be a platform for visible-near-IR entangled photon applications as it is at telecom wavelengths.

\section{Device design and fabrication}
The periodically poled lithium niobate waveguides (Figure~\ref{fig:design}a) were simulated in Lumerical MODE to determine the quasi-phase matching poling period. The guided modes at the design pump wavelength (406~nm) and SPDC center wavelength (812~nm) were simulated using the bulk Sellmeier coefficients of lithium niobate\cite{gayer_temperature_2008} and silicon dioxide\cite{toyoda_temperature_1983} with the geometric parameters shown in Figure~\ref{fig:design}c and a 60\degree sidewall, which is consistent with the fabrication process. To take advantage of lithium niobate’s largest nonlinear tensor element ($d_{33}$~=~28~pm/V)\cite{shoji_absolute_1997}, only the fundamental quasi-transverse electric (TE) modes were considered. An etch depth of 420~nm and a top width of 1.5~$\mu$m with a total LN thickness of 600~nm were targeted for ease of optical coupling, fabrication, and $\geq2~\mu$m poling period while providing high performance. For these parameters, the effective refractive indices ($n_{\text{eff,pump}}$~=~2.29, $n_{\text{eff,SPDC}}$~=~2.09) result in a quasi-phase matching poling period of $\Lambda=\lambda_{\text{pump}}/\Delta n_{\text{eff}}=2.03~\mu \text{m}$ at the target pump wavelength of 406~nm (Figure~\ref{fig:design}d).

\begin{figure}[b]
\centering\includegraphics[width=8.4cm]{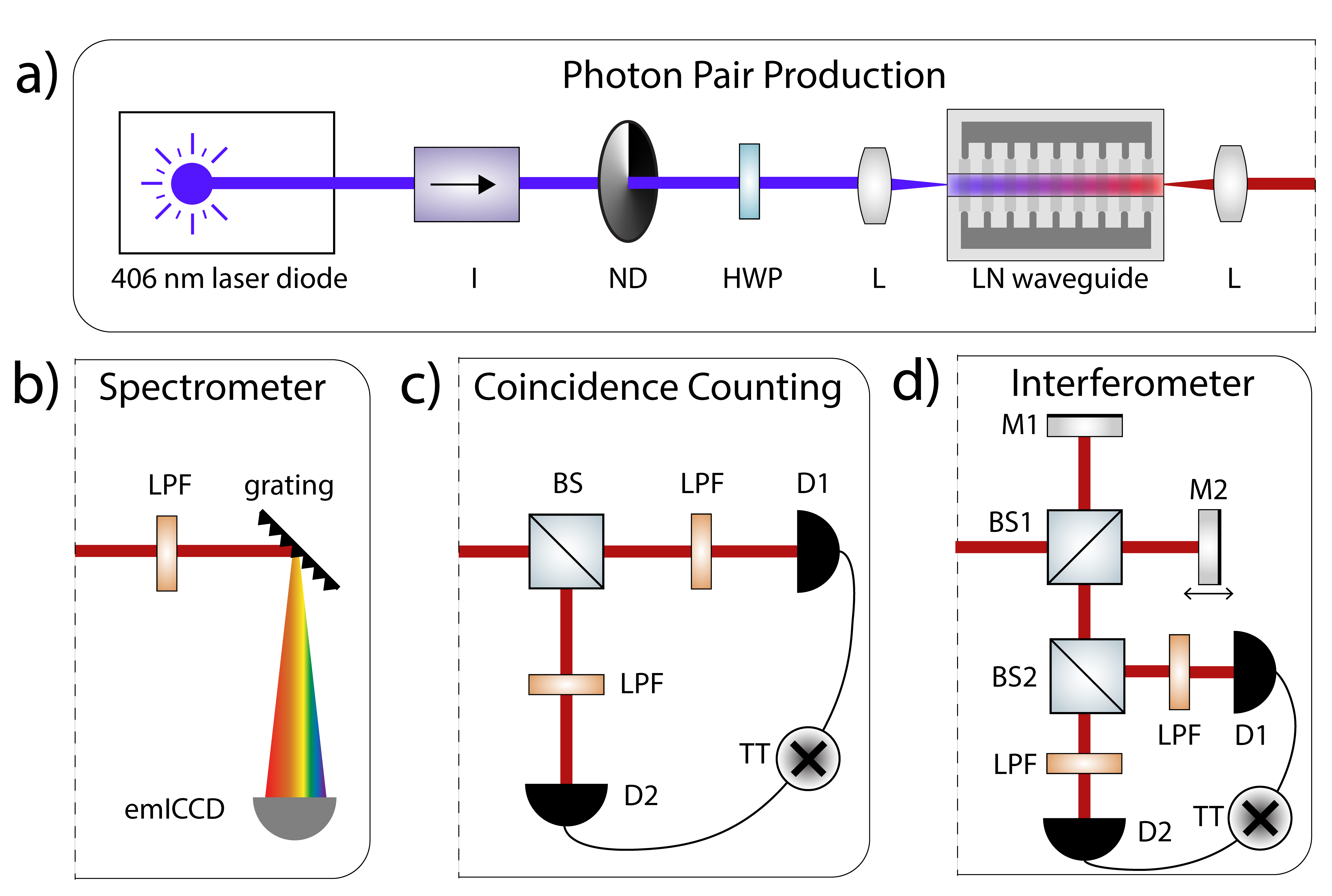}
\caption{\label{fig:setup} Schematic for coupling and detection of the photon pairs. I, isolator; ND, variable neutral density filter; HWP, half-wave plate; L, aspheric lens; LPF, long-pass filter; emICCD, electron-multiplying intensified charge-coupled device; BS, beamsplitter; D, single-photon avalanche detector; TT, time-tagger; M, mirror. (a)~Optical setup to couple into the TFLN waveguide. (b)~Characterization scheme to measure the photon pair spectra. (c)~Characterization scheme for coincidence counting. (d)~Optical setup for the Michelson interferometer.}
\end{figure}

The devices were fabricated from a 5\% MgO-doped X-cut thin-film lithium niobate on insulator wafer (NanoLN), which consists of 600~nm of lithium niobate bonded to 2~$\mu$m of silicon dioxide on a 0.4~mm silicon substrate. To quasi-phase match the SPDC sources, poling electrodes (7~mm long) were first fabricated by performing a metal lift-off through electron beam lithography with bilayer polymethyl methacrylate (PMMA) resist followed by electron beam evaporation of titanium (15~nm) and gold (55~nm). The electrodes were poled with a 490~V and 70~$\mu$s square wave, and the poled domain formation was monitored with second harmonic microscopy\cite{rusing_second_2019} (Figure~\ref{fig:design}b). After poling, waveguides were defined through an aligned electron beam lithography step with hydrogen silesoxquiane (HSQ) resist followed by argon inductive coupled plasma reactive ion etching to achieve an etch depth of 420~nm, verified through atomic force microscopy. The chip facets were manually polished to increase coupling efficiency. A broadband oscillator was used to verify the phase matching wavelength through second harmonic generation and was found to match with the computationally predicted second harmonic wavelength within $\pm$3~nm. This small discrepancy was likely due to fabrication tolerances, particularly in the etch depth and film thickness.

\section{Device characterization}
The spectrum, generation rate, and coherence properties of the entangled photon pairs produced from the fabricated device are characterized as shown in Figure~\ref{fig:setup}. In these experiments, the room-temperature periodically poled waveguide is pumped with a free-running laser diode (Coherent OBIS LX 405 nm) to produce entangled pairs (Figure~\ref{fig:setup}a). An antireflection-coated aspheric lens (NA = 0.58, Thorlabs C140TMD-A) couples the free-space pump beam to the fundamental TE mode of the waveguide. The photon pairs produced in the fundamental TE mode are collected off-chip and collimated using a similar aspheric lens (NA = 0.58, Thorlabs C140TMD-B).

\begin{figure}[b]
\centering\includegraphics[width=13cm]{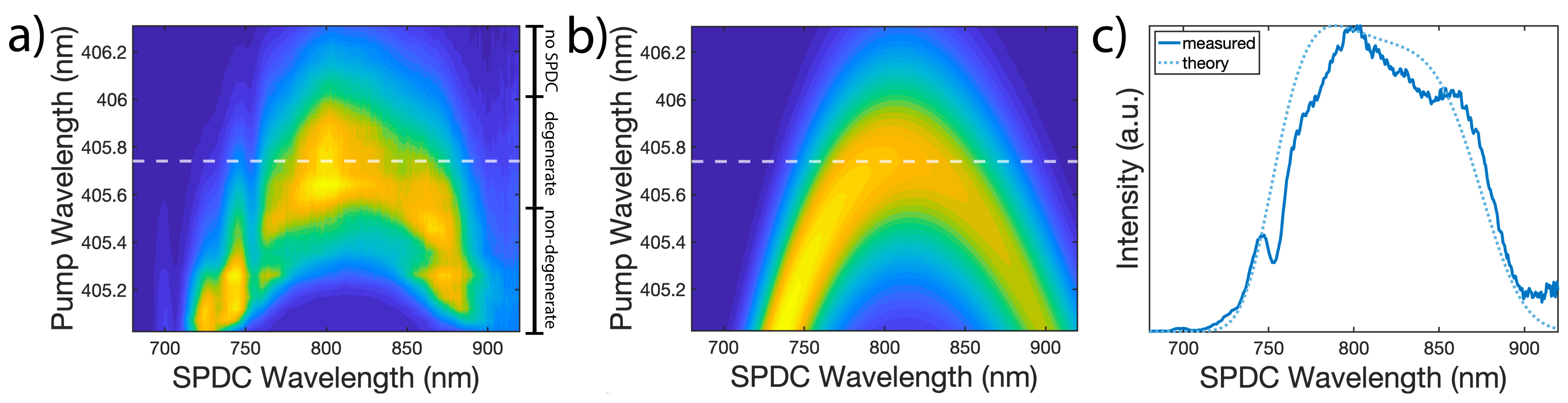} 
\caption{\label{fig:spectra} (a)~Measured and (b)~theoretical SPDC spectra as a function of pump wavelength. (c)~Lineouts of the measured and theoretical SPDC spectra at a pump wavelength of 405.7~nm, corresponding to the dashed lines in (a) and (b).}
\end{figure}

The spectra of the entangled photon pairs are measured to assess the phase-matching properties and tunability of the device (Figure~\ref{fig:spectra}). Pairs collected from the waveguide are transmitted to a grating spectrometer and measured using an electron-multiplying intensified camera (Figure~\ref{fig:setup}b). To tune the SPDC emission, the center wavelength of the pump wavelength is varied from 405-406.4~nm by changing the drive current of the laser diode. Variable neutral-density filters are used to keep the pump power for all the collected spectra at a consistent 10~$\mu$W. In doing so, three distinct phase matching regions are explored (Figure~\ref{fig:spectra}a): 1) at long pump wavelengths, the phase matching condition is not satisfied and emission is not observed; 2) from 405.6-405.9 nm, the degenerate wavelengths are phasematched; and 3) at short pump wavelengths, the spectrum splits and nondegenerate emission extending to the cutoff wavelength of the filter is observed. The dip in intensity in Figure~\ref{fig:spectra}a around 760~nm is likely due to impurities in the thin film, contamination during the device fabrication process, or ambient absorption. Due to the linewidth of the laser used in the experiment (0.8~nm FWHM), the spectra are considerably broadened compared to the spectra expected from a single-frequency pump laser (Supplemental Figure S3). Nevertheless, experiment and theory reach qualitative agreement by factoring the pump linewidth into the calculations (Figure~\ref{fig:spectra}b). For all subsequent experiments, a laser center wavelength of 405.7~nm is used for degenerate phase matching. The resulting spectrum (Figure~\ref{fig:spectra}c) is centered at 811.4 $\pm$ 0.7~nm with a FWHM bandwidth of 117 nm (53 THz) that accounts for 85\% of the overall flux. 

\begin{figure}[t]
\centering\includegraphics[width=13cm]{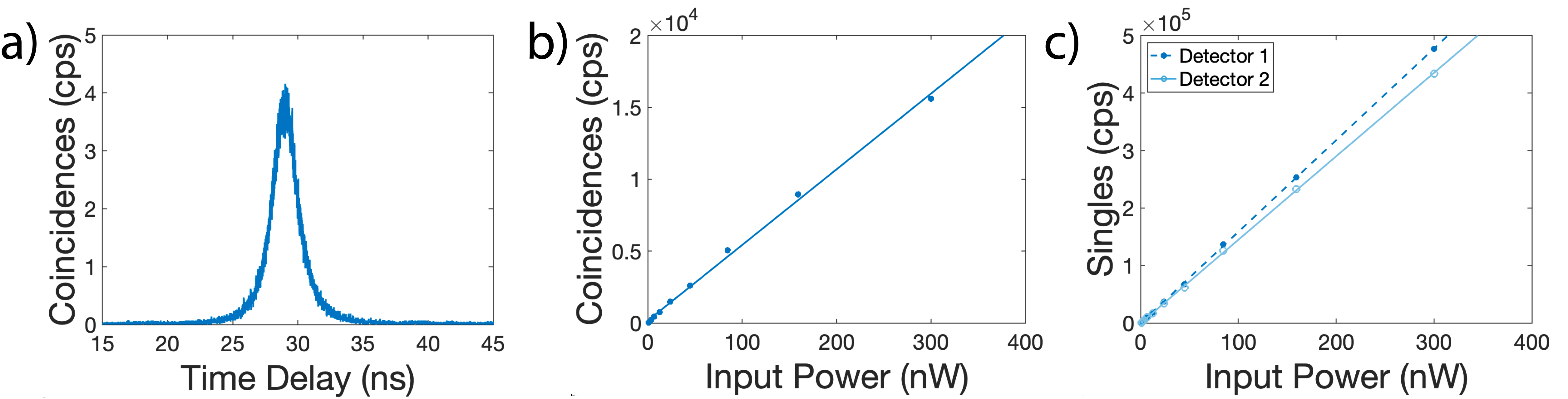} 
\caption{\label{fig:coincidence} (a)~Coincidence histogram at an input pump power of 45~nW including accidentals. Measured (b)~coincidence and (c)~singles counts while sweeping the input power. The fitted slopes (solid lines) produce the pair generation efficiency of our device.}
\end{figure}

\begin{figure}[bt!]
\centering\includegraphics[width=13cm]{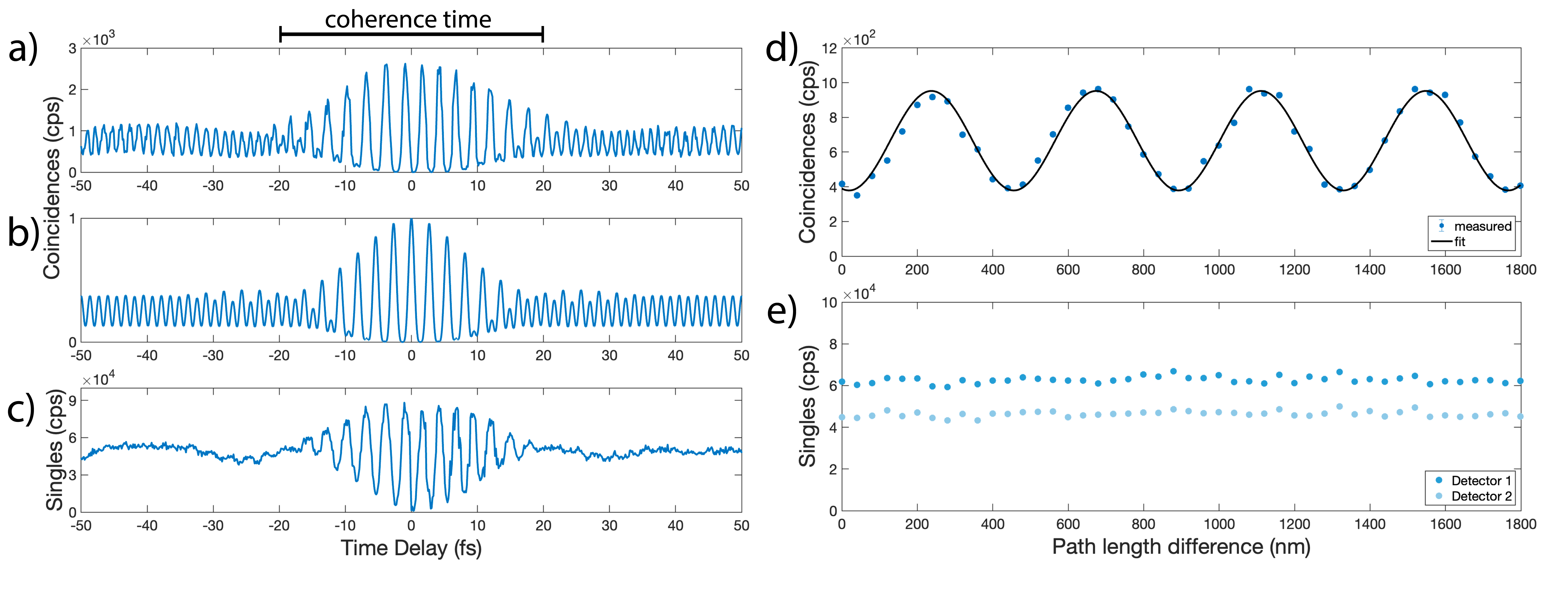}  
\caption{\label{fig:michelson} The (a)~measured and (b)~simulated (Supplemental) two photon Michelson interferogram for the device. (c)~Corresponding singles counts out of the interferometer, demonstrating single photon interference within the coherence length of the source. The (d)~coincidence counts and (e)~singles counts far from the coherence length of the source. Note that all measured coincidence counts ((a) and (d)) do not have accidentals subtracted.}
\end{figure}

The pair generation efficiency of the device is measured through coincidence counting between two SPADs (Figure~\ref{fig:setup}c). In this experiment, the pairs from the chip were split at a 50:50 broadband beamsplitter (Thorlabs BS014) and coupled to multimode optical fibers connected to the detectors. Coincidence detection events between the SPADs were recorded with a time-tagger (Picoquant PicoHarp 300). Figure~\ref{fig:coincidence}a shows a representative coincidence histogram recorded at 45~nW of pump power. The temporal correlation in this graph (3.4~ns FWHM) is given by the response time of the SPADs and not the entangled photon correlations (see Figure~\ref{fig:michelson} later in the text). The coincidence counts are corrected by background subtraction of the number of counts in a 9.5~ns window at the histogram peak from the number of counts in another 9.5~ns window in a background region far from the peak. Sweeping the laser power with a neutral density filter yields the curves in Figure~\ref{fig:coincidence}b-c, which are linearly fit to determine the pair generation efficiency. To account for the wavelength dependence of the SPAD quantum efficiency (Supplemental Equation S11), all wavelengths in the spectrum are integrated over to calculate the average detection efficiency for single photons ($\eta_{1}=0.52$) as well as the average joint pair detection probability ($\eta_{12}=0.27$). Including a factor of 2 due to the probability of splitting pairs at the beamsplitter yields Equation~\ref{eq:efficiency} for the measured efficiency of the source:

\begin{equation} \label{eq:efficiency}
    E = \frac{m_1 m_2}{m_c} \frac{\eta_{12}}{2 \eta_1^2}
\end{equation}
Here $E$ is the pair generation efficiency, $m_1$ and $m_2$ are the singles rates at the two detectors, and $m_c$ is the rate of coincidences, all in units of counts/mW. Accounting for the 10.2 dB transmission loss of the pump laser into the waveguide, a pair generation efficiency of $(2.3\pm 0.5) \times 10^{11}$ pairs/mW is measured, which is equivalent to a per-pump-photon efficiency of $(1.1\pm0.2) \times 10^{-4}$ pairs/photon. Over the 117~nm FWHM bandwidth of the spectrum, this efficiency translates to an average brightness of $(1.6 \pm 0.3) \times 10^{9}$ pairs/nm/mW. The uncertainties here and throughout this work are reported as one standard deviation, derived from the standard error in the fits of Figure~\ref{fig:coincidence}b-c and the uncertainty in the detector quantum efficiency. The ratio of singles counts to coincidence counts suggests that the transmission of the SPDC from the waveguide to each of the two detectors is 8.4~dB and 8.0~dB, respectively, which includes losses out of the waveguide and of the free-space optics. Our theoretical efficiency (Supplemental), including the FWHM of the pump laser, is $2.66 \times 10^{11}$~pairs/mW, in close agreement with our experimental results. 

Finally, the two-photon interference is measured (Figure~\ref{fig:setup}d) to demonstrate the non-classical behaviour of the produced photon pairs. Figure~\ref{fig:michelson} shows the measured two-photon interferogram (Figure~\ref{fig:michelson}a) obtained from the device without subtracting accidentals, as well as the one-photon interferogram (Figure~\ref{fig:michelson}c) for comparison. Due to the aforementioned temporal resolution of the SPADs used here, the four unique paths through the interferometer are indistinguishable and combine to yield the interference pattern. The important features of the interferogram are as follows: 1) Near the zero path length difference, photons are delayed within the coherence length of the source and exhibit both one- and two-photon interference. A visibility of 100.$\pm$1\% is measured within the coherence length, with an uncertainty derived assuming Poissonian statistics.
This near-perfect visibility indicates good mode overlap in the interferometer and indistinguishability of the photons within the pair. 2) Far from the coherence length of the source (delays greater than 20 fs), interference between two photons taking different paths disappears, which explains why the single-photon interference disappears in Figure~\ref{fig:michelson}c and \ref{fig:michelson}e. Notably, interference between pairs of photons that travel together through the interferometer persists with a fringe period at half the pair wavelength and is shown in more detail in Figure~\ref{fig:michelson}d. This feature suggests quantum interference due to the energy-time entanglement of the pairs, and would not be observed if the light was generated from a coherent or a thermal source with a similar spectrum\cite{ou_michelson_1990}. A fringe visibility of $43\pm3\%$ is observed in this region far from the coherence length, which is close to the theoretical maximum of 50\% for this experiment due to the temporal resolution of the detectors. The uncertainty in visibility is given from the standard error in the fit of Figure~\ref{fig:michelson}d. The qualitative agreement between the measured and theoretical two-photon interferogram (Figure~\ref{fig:michelson}a-b) suggests that SPDC and genuine energy-time entanglement are being produced.

\section{Discussion}
\begin{figure}[bt!]
\centering\includegraphics[width=7cm]{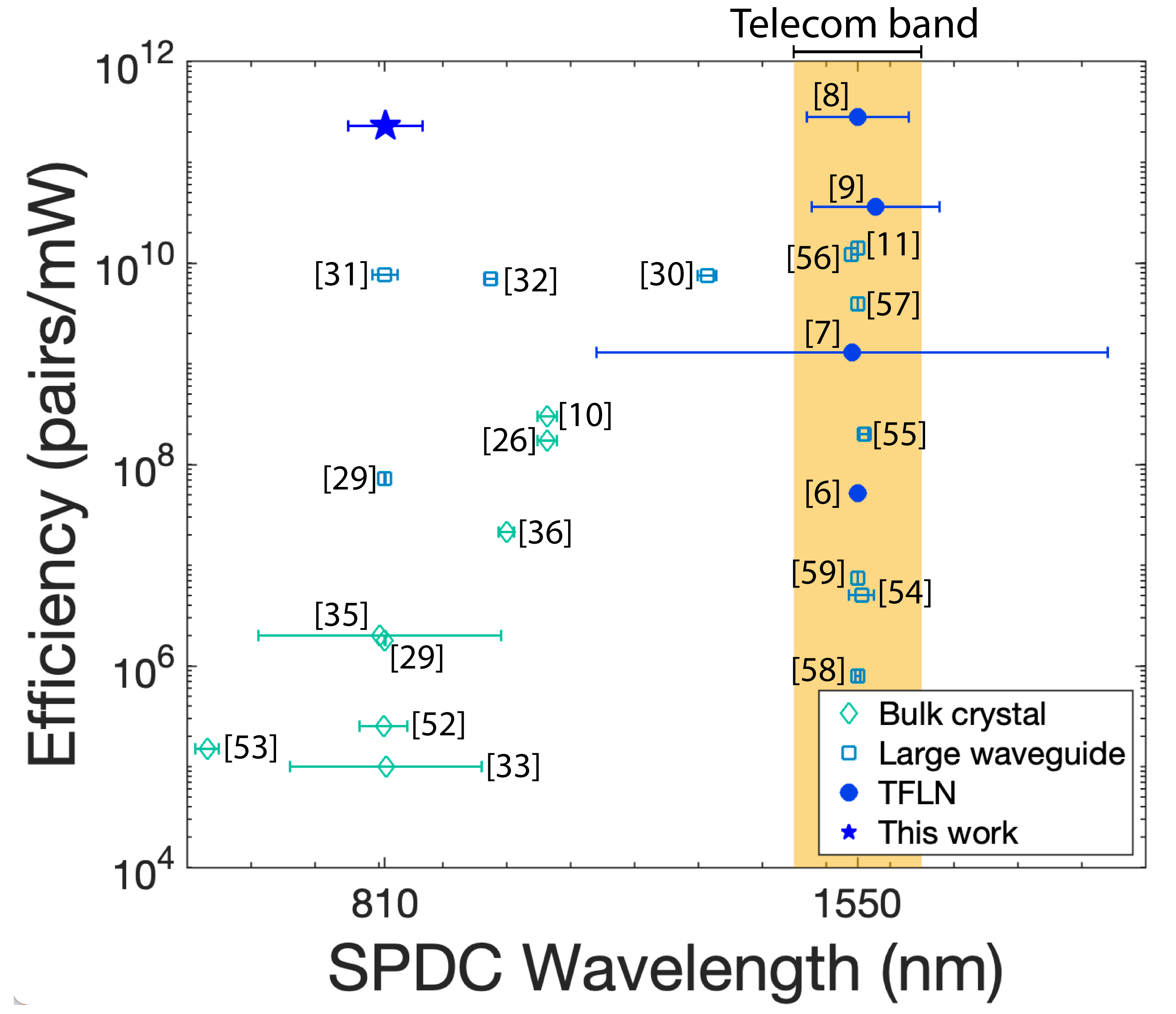}
\caption{\label{fig:literature} Comparison of relevant literature SPDC sources to this work. Horizontal error bars represent the reported bandwidth of the sources. Data for the efficiency, center wavelength, and bandwidth are taken from Refs.~\cite{fiorentino_spontaneous_2007,nasr_ultrabroadband_2008,okano_054_2015,okano_generation_2012,sensarn_generation_2010,lavoie_phasemodulated_2020,tabakaev_spatial_2022,dayan_nonlinear_2005} for bulk crystal sources, Refs.~\cite{fiorentino_spontaneous_2007,tanzilli_highly_2001,jechow_high_2008,gabler_photon_2023,kang_monolithic_2016,jin_-chip_2014,meyer-scott_certifying_2016,ding_hybrid-cascaded_2015,bock_highly_2016,sarrafi_high-visibility_2014,fujii_bright_2007} for large waveguide sources (including micromachined and diffused waveguides), and Refs.~\cite{xue_ultrabright_2021,javid_ultrabroadband_2021,zhao_high_2020,zhang_scalable_2023} for TFLN sources. The yellow shaded region represents the typical telecommunication wavelength window.}
\end{figure}

Compared to the state of the art for visible photon pair sources, the device presented in this work exhibits substantially improved brightness and efficiency due to the small effective area of the waveguide. The device's performance against reported literature values is plotted in Figure~\ref{fig:literature} using efficiency and wavelength as the figures of merit. Telecom and MIR TFLN devices are also included, demonstrating comparable or improved performance. At short wavelengths, the phase-matching bandwidth decreases due to group velocity dispersion in lithium niobate, but the efficiency remains high because the downconversion spectral power density scales with the inverse fifth power of the SPDC wavelength ($\lambda_s^{-5}$)\cite{fiorentino_spontaneous_2007}. In addition to the benefits of higher-energy photon pairs, the strategy of decreasing the SPDC wavelength to access higher efficiencies could enable single-photon nonlinearities when integrated with a resonator\cite{Jankowski_dispersion_2021}. The high efficiency of this device, which is limited here by the linewidth of the pump laser, has significant implications for practical uses of entangled photons, including reducing integration times, allowing the use of low-power laser diodes for pair generation, and reducing fluorescence and stray light. Even with the losses present in this setup, high signal to noise coincidence measurements are acquired with nanowatts of laser power, highlighting the potential of this device.

The efficiency and brightness of the device can be further improved by narrowing the bandwidth of the pump laser, as discussed further in the Supplemental. A single-frequency pump is estimated to shrink the phase-matching bandwidth from 117 nm to 35 nm, increasing the brightness by a comparable factor. Furthermore, the SPDC process is most efficient near degeneracy because the group velocity of the signal and idler are equal to first order, so the efficiency could potentially increase by a factor of 5. However, one benefit of using a multifrequency pump is that the device sensitivity to the laser wavelength is reduced, yielding a higher stability average response with greater bandwidth. 

\section{Conclusion}
Efficient photon pair generation has been demonstrated with an integrated thin-film lithium niobate waveguide at visible and NIR wavelengths (720-900~nm). An on-chip SPDC efficiency of $(2.3\pm 0.5) \times 10^{11}$ pairs/mW, which is on-par with reported literature at telecom wavelengths, has been produced near the usually associated cut-off wavelengths for the pump (406~nm) in TFLN. The photon pair spectra has an average brightness of $(1.6 \pm 0.3) \times 10^{9}$ pairs/mW/nm, centered at 811~nm with a 117~nm bandwidth. To date, these results are the shortest wavelength photon pairs generated in a thin film platform by nearly an octave. The work opens up opportunities to exploit the quantum advantage of integrated entangled photon circuits beyond telecom to imaging and spectroscopy applications. 

\begin{backmatter}
\bmsection{Funding}
Air Force Office of Scientific Research (FA9550-20-1-0040); Army Research Office (W911NF-23-1-0048); National Science Foundation (EECS 1846273, DGE-1745301); U.S. Department of Energy (DE-SC0020151).

\bmsection{Acknowledgments}
The authors gratefully acknowledge the critical support and infrastructure provided for this work by The Kavli Nanoscience Institute (KNI) and the Beckman Biological Imaging Facility at Caltech. This work was additionally supported by the KNI-Wheatley Scholar in Nanoscience and the Rothenberg Innovation Initiative. N.H. was supported by the Department of Defense (DoD) through the National Defense Science and Engineering Graduate (NDSEG) Fellowship Program. E.H. was supported by the National Science Foundation Graduate Research Fellowship Program under Grant no. DGE‐1745301. Any opinion, findings, and conclusions or recommendations expressed in this material are those of the authors(s) and do not necessarily reflect the views of the National Science Foundation. 

\bmsection{Disclosures}
The authors declare no conflicts of interest.

\bmsection{Data Availability Statement}
Data underlying the results presented in this paper are not publicly available at this time but may be obtained from the authors upon reasonable request.

\bmsection{Supplemental document}
See Supplement 1 for supporting content. 

\end{backmatter}


\bibliography{main}

\end{document}


\maketitle

\section{SPDC Theoretical Calculations}

The theory of parametric down-conversion in waveguides is well established. Here, the equations used in the simulations of the main text are reviewed. We consider a strong, undepleted pump wave at frequency $\omega_p$, stimulating the production of a signal photon with frequency $\omega_s$ and idler photon with frequency $\omega_i=(\omega_p-\omega_s)$. From Equation \ref{eq:spectraldensity}, the rate of downconversion $W$ is given by\cite{fiorentino_spontaneous_2007}:

\begin{equation}
\frac{\textrm{d}W}{\textrm{d}\omega_s} =
\frac{d_{\text{eff}}^2 P_p \omega_s \omega_i}{\pi \varepsilon_0 c^3 n_p n_s n_i}
\frac{L^2}{A_{\text{eff}}}
\textrm{sinc}^2 \left( \frac{\Delta k L}{2} \right)
\label{eq:spectraldensity}
\end{equation}
where $d_{\text{eff}}$ is the effective nonlinear coefficient of the interaction, $P_p$ is the power of the pump, $\varepsilon_0$ is the vacuum permittivity, $c$ is the speed of light, and $n_{p,s,i}$ are the respective effective refractive of the modes. The momentum-mismatch $\Delta k$ is

\begin{equation}
\Delta k = k_p - k_s - k_i - k_{QPM}
\label{eq:phasematching}
\end{equation}
where $k_{p,s,i}$ are the respective propagation constants of the modes and $k_{QPM}$ is the wavevector due to quasi-phase-matching.  $k_{QPM}$ and $d_{\text{eff}}$ both take the order of the phase-matching $m$ into account, which is unity for the device used in this work:

\begin{equation}
k_{QPM} = \frac{2 \pi m}{\Lambda}
\label{eq:polingperiod}
\end{equation}

\begin{equation}
d_{\text{eff}} = \frac{2 d}{\pi m}
\label{eq:effectivenonlinearity}
\end{equation}
Here $\Lambda$ is the poling period and $d$ is the nonlinear tensor element of the interaction. The largest nonlinear tensor element of lithium niobate is the $d_{33}$ element; the remaining elements are approximately an order of magnitude or more smaller. To maximize the SPDC efficiency in this work, we utilized $d_{33} = 28.4$~pm/V (for 5\% MgO-doped lithium niobate at 852~nm)\cite{shoji_absolute_1997} to obtain $d_{\text{eff}} = 18.1$~pm/V through Equation~\ref{eq:effectivenonlinearity}.

In the waveguide, the effective area $A_{\text{eff}}$ is given as a mode overlap calculation of the three interacting waves mediated through the nonlinear tensor of the crystal\cite{jankowski_ultrabroadband_2020}. For the device used in this work, the effective area can be simplified as:

\begin{equation}
A_{\text{eff}} = \frac{A_{p}A_{s}A_{i}}{\displaystyle \sqrt{\frac{n_p n_s n_i A_p A_s A_i}{8 Z_{0}^{3} P^{3}}} \left| \iint \overline{d}_{33} E^{*}_{s}E^{*}_{i}E_{p}\,dx\,dy\right|^{2}} 
\label{eq:effectivearea}
\end{equation}
Here $A_{p,s,i}$ are the effective areas of each mode, $Z_{0}$ is the free space impedance, $P$ is the power in each mode (normalized to 1~W here), $\overline{d}_{33}$ is the normalized $d_{33}$ nonlinear tensor element, and $E_{p,s,i}$ are the electric field profiles of each mode. The mode profiles are evaluated in Lumerical MODE to find that $A_{\text{eff}} = 0.7~\mu\textrm{m}^2$.

To modify \ref{eq:spectraldensity} to accommodate a multi-frequency pump, the pump power $P_p$ can be replaced with the spectral intensity of the pump $P(\omega_p)$ and integrated over all frequencies:

\begin{equation}
\frac{\textrm{d}W}{\textrm{d}\omega_s} =
\int 
\frac{d_{\text{eff}}^2 \omega_s \omega_i}{\pi \varepsilon_0 c^3 n_p n_s n_i}
\frac{L^2}{A_{\text{eff}}}
\textrm{sinc}^2 \left( \frac{\Delta k L}{2} \right)
P(\omega_p) \textrm{d} \omega_p
\label{eq:multimodespectraldensity}
\end{equation}
For the simulations given in the main text, the pump spectral intensity is approximated as a Gaussian distribution (Equation \ref{eq:gaussian}) with a mean $\mu$ and standard deviation $\sigma$ such that the FWHM is 0.8~nm, as measured experimentally, and is normalized to an total intensity of 1~mW.

\begin{equation}
P(\omega_p) = \frac{1}{\sigma \sqrt{2 \pi}} 
\textrm{exp} \left[ -\frac{1}{2} \left( \frac{\omega_p-\mu}{\sigma} \right)^2 \right]
\label{eq:gaussian}
\end{equation}
Because the signal and idler are indistinguishable, and thus both detected in our experiment, the expected total number of counts at a given frequency for the spectrally resolved experiments (Figure~3 of the main text) of is:

\begin{equation}
\frac{\textrm{d}W}{\textrm{d}\omega} =
\left. \frac{\textrm{d}W}{\textrm{d}\omega} \right|_{\omega = \omega_s} + 
\left. \frac{\textrm{d}W}{\textrm{d}\omega} \right|_{\omega = \omega_p - \omega_s}
\label{eq:totcounts}
\end{equation}

Finally, it is useful to have an equation for the probability of jointly detecting a signal photon at $\omega_s$ and an idler at $\omega_i$. This distribution is typically called the Joint Spectral Intensity (JSI) and is obtained from \ref{eq:multimodespectraldensity} by replacing the pump frequency by sum of the signal and idler frequencies:

\begin{equation}
\textrm{JSI}(\omega_s, \omega_i) =
C \frac{\omega_s \omega_i}{n_p n_s n_i}
\textrm{sinc}^2 \left( \frac{\Delta k L}{2} \right)
P(\omega_s + \omega_i)
\label{eq:jsi}
\end{equation}
Here, $C$ is a normalization constant chosen such that 

\begin{equation}
\iint \textrm{JSI}(\omega_s, \omega_i) \textrm{d}\omega_s \textrm{d}\omega_i \equiv 1
\label{eq:jsinorm}
\end{equation}
Thus, the JSI represents a two-dimensional probability distribution, and is symmetric with respect to exchange of the variables $\omega_s$ and $\omega_i$. 

\section{Single and Joint Detection probability}

\begin{figure}[b]
\centering
{\includegraphics[width=.6\linewidth]{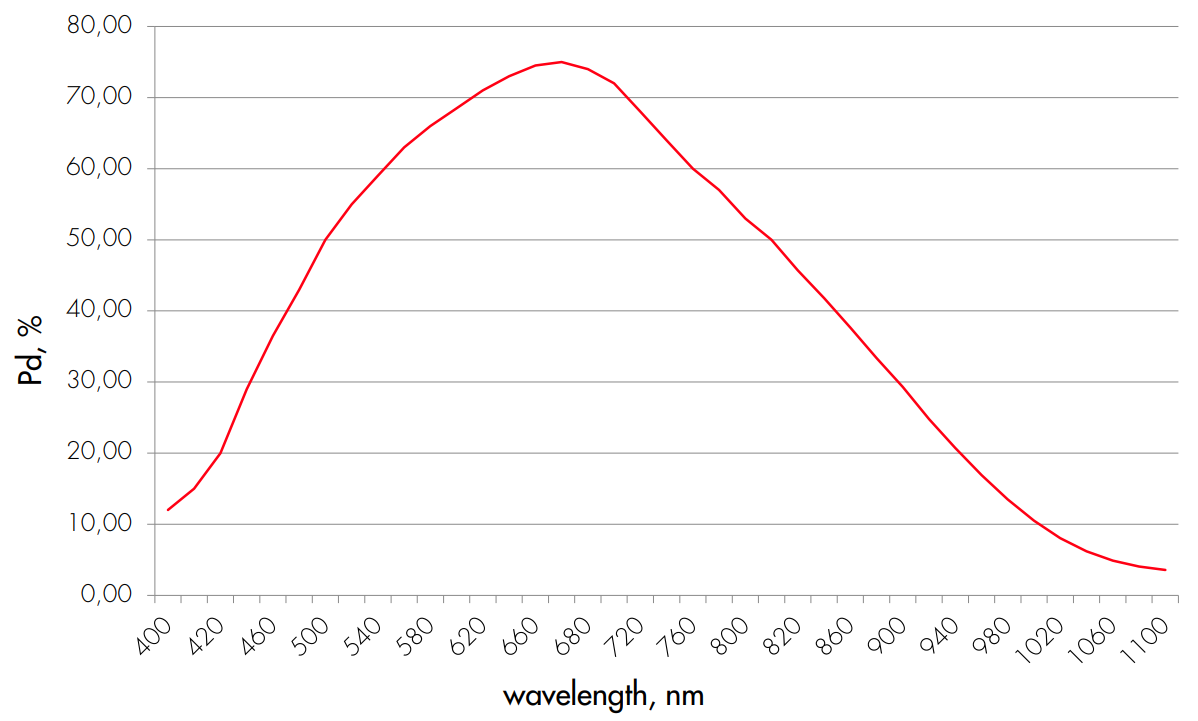}}
\caption{Spectrally-dependent quantum efficiency $\textrm{QE}(\omega)$, adapted from \cite{laser_components_single_nodate}.}
\label{fig:QE}
\end{figure}

The SPADs used in this study (Laser Components COUNT-10C) exhibit a spectrally-dependent quantum efficiency with a peak at 680~nm (Figure~\ref{fig:QE}). This wavelength dependence complicates the prediction of detection probabilities due to the large SPDC bandwidth.

The quantum efficiency curve from the manufacturer specifications can be interpolated to obtain a continuous quantum efficiency $\textrm{QE}(\omega)$, which ranges from 0 to 1. From this distribution, the probability of detecting both the signal and idler (i.e. a coincidence) is an expectation value given as:
\begin{equation}
\eta_{12} = 
\iint \textrm{JSI}(\omega_s, \omega_i) 
\textrm{QE}(\omega_s) \textrm{QE}(\omega_i)
\textrm{d}\omega_s \textrm{d}\omega_i
\label{eq:coincidenceprob}
\end{equation}
The probability of detection given the presence of a single photon is also given as an expectation value. Here, we use the quantum efficiency weighted by the probability distribution of the signal photon frequency, which is obtained by integrating over all possible idler frequencies:
\begin{equation}
\eta_{1} = 
\iint \textrm{JSI}(\omega_s, \omega_i) 
\textrm{QE}(\omega_s)
\textrm{d}\omega_s \textrm{d}\omega_i
\label{eq:singleprob}
\end{equation}

These probabilities are calculated as $\eta_{1}=0.522$ and $\eta_{12}=0.266$. In this case, $\eta_{12}\approx \eta_1^2$, which would not be generally expected given the wide range of wavelengths and large variations in the quantum efficiency. However, due to the linearity of $\textrm{QE}(\omega)$ in the vicinity of our experiment, an increase in the quantum efficiency of the signal photon at shorter wavelengths is balanced by the decreased probability of detecting the idler photon at longer wavelengths.

\section{Power Scaling Equations}

The scaling of the single and coincidence detection rates are useful for estimating the device efficiency and transmission efficiency of pairs through the system. A simple model is used to estimate these quantities, similar to Ref.~\cite{tanzilli_highly_2001}. First, the rate of singles counts at detectors 1 and 2 ($S_1$, $S_2$) are modeled as:

\begin{equation}
S_1(P) = 2 E P \mu_1 R \eta_1 + \textrm{Dark}_1
\label{eq:singles_det1}
\end{equation}

\begin{equation}
S_2(P) = 2 E P \mu_2 T \eta_1 + \textrm{Dark}_2
\label{eq:singles_det2}
\end{equation}
where $E$ is the pair production efficiency, $P$ is the on-chip power, $\mu_{1,2}$ are the transmission of the path to detectors 1 and 2, $R$ and $T$ are the beamsplitter reflection and transmission probabilities, $\eta_1$ is the average detector quantum efficiency given by Equation~\ref{eq:singleprob}, and $\textrm{Dark}_{1,2}$ are the dark counts of detectors 1 and 2. This can be interpreted as $2EP$ photons being produced by the device, then suffering losses of $\mu_{1}R\eta_1$ or $\mu_2T\eta_1$ in the setup before being detected by the respective SPADs.

Similarly, the rate of coincidences ($S_{cc}(P)$, which requires the detection of both photons, is:

\begin{equation}
S_{cc}(P) = E P \mu_1 \mu_2 2 R T \eta_12
\label{eq:singles_cc}
\end{equation}
where the factor of $2RT$ accounts for the two possible paths through the beamsplitter that result in a coincidence.

The slopes of Equations~\ref{eq:singles_det1}-\ref{eq:singles_cc} with respect to power, as shown below in Equations~\ref{eq:singles_det1_slope}-\ref{eq:cc_slope}, can be experimentally measured:

\begin{equation}
m_1 = 2 E \mu_1 R \eta_1
\label{eq:singles_det1_slope}
\end{equation}

\begin{equation}
m_2 = 2 E \mu_1 R \eta_1
\label{eq:singles_det2_slope}
\end{equation}

\begin{equation}
m_{cc} = 2 E \mu_1 \mu_2 R T \eta_{12}
\label{eq:cc_slope}
\end{equation}

These equations are then rearranged to solve for the efficiency and loss of the setup, given that $R$, $T$, $\eta_1$, and $\eta_{12}$ are known.

\begin{equation}
E = \frac{m_1 m_2}{m_{cc}}\frac{\eta_{12}}{2\eta_1^2}
\label{eq:efficiency}
\end{equation}

\begin{equation}
\mu_1 = \frac{m_{cc}}{m_2}\frac{\eta_1}{R\eta_{12}}
\label{eq:loss_1}
\end{equation}

\begin{equation}
\mu_2 = \frac{m_{cc}}{m_1}\frac{\eta_1}{T\eta_{12}}
\label{eq:loss_2}
\end{equation}

The parameters measured in the experiment as well as the parameters necessary for calculating the device loss are displayed in Table~\ref{tab:parameters}. 

\begin{table}[tb]
\centering
\caption{\textbf{Power Scaling Parameters.} Uncertainties of the photon count rates are obtained from standard error of a linear fit. Beamsplitter and detector uncertainties are obtained from the manufacturers. Derived parameter ($E,\mu_1,\mu_2$) uncertainties  are obtained through standard uncertainty propagation.}
\begin{tabular}{ccc}
\hline
Parameter & Value & Description  \\
\hline
$m_1$ & $(1.670 \pm 0.005) \times 10^{10} ~\textrm{Hz/mW}$ & Detector 1 Singles Rate \\
$m_2$ & $(1.521 \pm 0.007) \times 10^{10} ~\textrm{Hz/mW}$ & Detector 2 Singles Rate \\
$m_{cc}$ & $(5.5 \pm 0.1) \times 10^8 ~\textrm{Hz/mW}$ & Coincidence Rate \\
$R$ & $47 \pm 10 \%$ & Beamsplitter Reflectance \\
$T$ & $47 \pm 10 \%$ & Beamsplitter Transmittance \\
$\eta_1$ & $52 \pm 5 \%$ & Average Quantum Efficiency \\
$\eta_{12}$ & $27 \pm 4 \%$ & Average Joint Quantum Efficiency \\
$E$ & $(2.3\pm 0.5) \times 10^{11}$ pairs/mW & Device Efficiency \\
$\mu_1$ & $15\pm 4 \%$ & Transmission to Detector 1 \\
$\mu_2$ & $14\pm 4 \%$ & Transmission to Detector 2 \\
\hline
\end{tabular}
  \label{tab:parameters}
\end{table}

\section{Michelson Interferogram}

\begin{figure}[b]
\centering
\includegraphics[width=13cm]{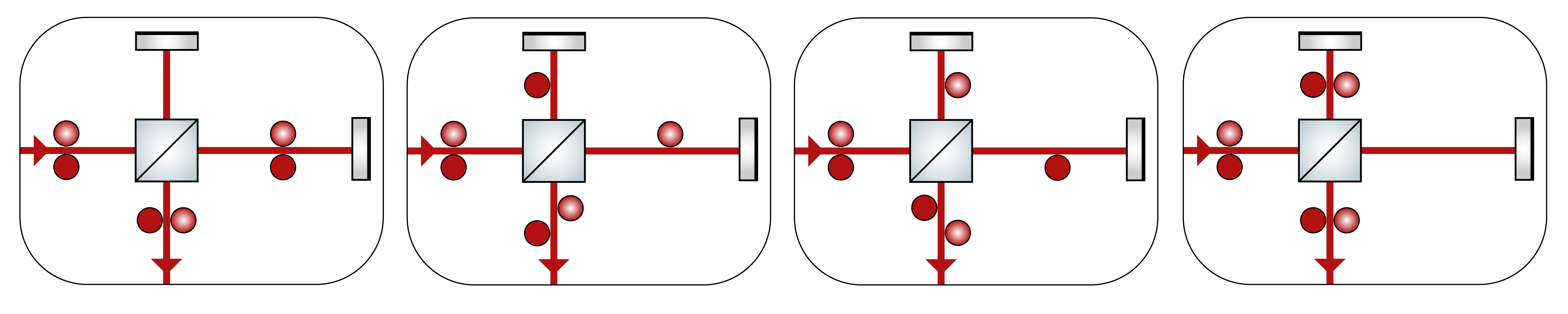}
\caption{Schematic of the four unique paths a photon pair can take through the Michelson interferometer.}
\label{fig:michelsonpath}
\end{figure}

Two-photon interference in a Michelson interferometer has been demonstrated in several previous works. Equation 16 from Ref.~\cite{lopez-mago_coherence_2012} is used here to predict the interferogram shown in Figure 5 of the main text.

\begin{equation}
R_{cc}(\tau) = \iint d\omega_s d\omega_p ~\textrm{JSI}(\omega_s, \omega_p-\omega_s)[1 + \textrm{cos}(\omega_s \tau)][1+\textrm{cos}((\omega_p-\omega_s)\tau)]
\label{eq:michelson}
\end{equation}

\begin{multline}\label{eq:cosines}
[1 + \textrm{cos}(\omega_s \tau)][1+\textrm{cos}((\omega_p-\omega_s)\tau)] = \\ 1 + \textrm{cos}(\omega_s \tau) + \textrm{cos}((\omega_p - \omega_s)\tau) + \frac{1}{2}\textrm{cos}(\omega_p \tau) + \frac{1}{2}\textrm{cos}((\omega_p-2\omega_s)\tau)
\end{multline}

Here $\tau$ is the path-length difference of the interferometer, which is twice the physical difference in the path lengths due to the reflection at the translating mirror. Equation \ref{eq:cosines} expands the cosine terms in \ref{eq:michelson}. At large delays, the cosine terms dependent on $\omega_s$ average to zero, and the interference is dominated by $1 + \textrm{cos}(\omega_p \tau)$, explaining the oscillation at the pump frequency with a theoretical 50\% visibility.

\section{Pump Linewidth Comparison}
The measured SPDC efficiency of the device ($(2.3\pm 0.5) \times 10^{11}$ pairs/mW) is on par with the efficiencies of telecom regime TFLN sources and an order of magnitude or greater than the efficiencies of other visible SPDC sources, as shown in Figure 6 of the main text. However, this efficiency is significantly lowered by the bandwidth of the pump laser used in this work, which was experimentally measured to be 0.8~nm FWHM. 

\begin{figure}[tb]
\centering
\includegraphics[width=12cm]{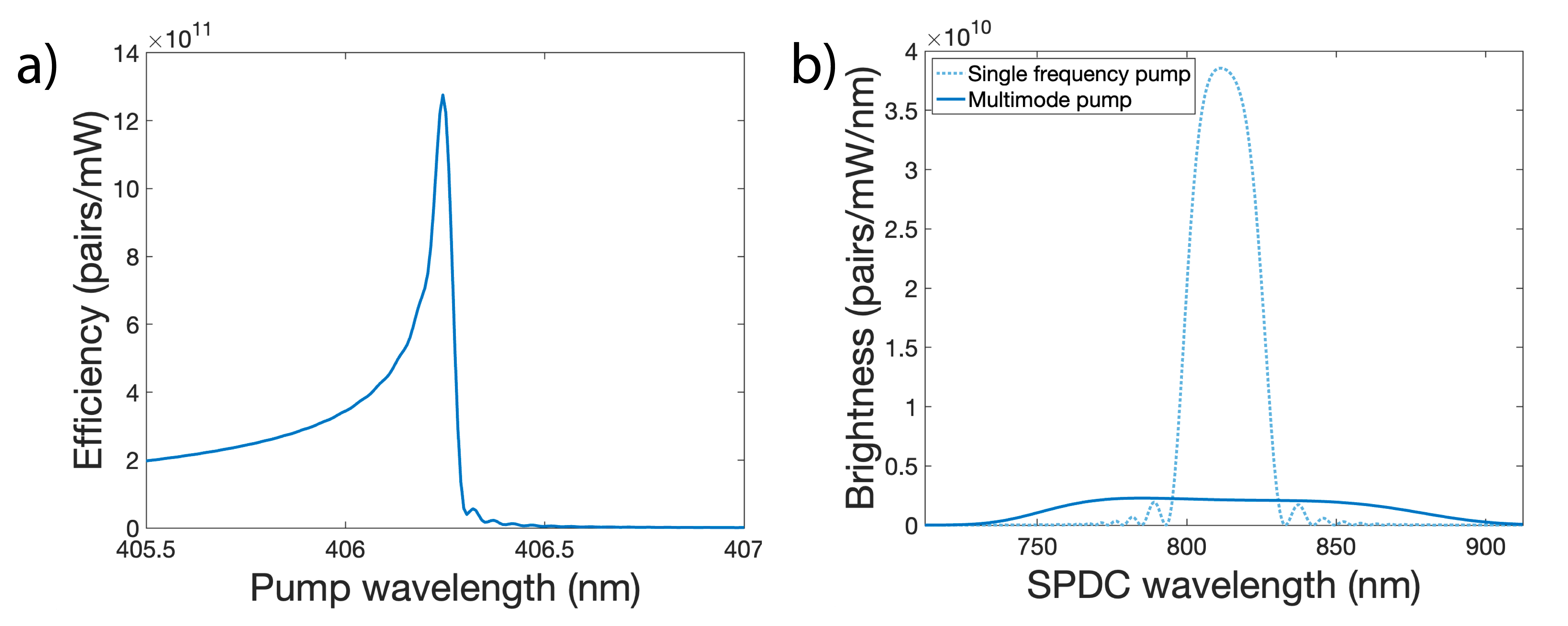}
\caption{\textbf{Comparison of pump linewidths.} (a)~Theoretical single-frequency SPDC efficiency while varying the pump wavelength. The lack of symmetry in the efficiency can be explained by non-degenerate phase-matching at wavelengths shorter than the degenerate wavelength (here designed to be 406.25~nm), as shown by Figure~3 of the main text. (b)~Theoretical SPDC spectra for a single frequency pump and a broadband pump, both with 1~mW of power. }
\label{fig:theoefficiency}
\end{figure}

As shown in Figure~\ref{fig:theoefficiency}a, the calculated SPDC efficiency for a single frequency pump demonstrates a strong peak at the degenerate wavelength. The SPDC from a multimode pump can be understood as a convolution of the pump lineshape with Figure~\ref{fig:theoefficiency}a. For the same total input power, the spread in the spectral density of a multimode pump results in only a fraction of the total power overlapping with this peak in the efficiency, while a single frequency pump can be tuned to the maximum efficiency. Therefore, as shown in Figure~\ref{fig:theoefficiency}b, the maximum brightness of the SPDC spectrum from a single frequency pump is an order of magnitude higher than the spectrum for a multimode pump for the same input power.

To further quantify the enhanced device performance with a single frequency pump, the theoretical efficiencies of SPDC from single frequency pump and a multimode pump can be calculated. Approximating the pump as a Gaussian distribution with a FWHM of 0.8~nm (Equation~\ref{eq:gaussian}), theoretical efficiency of $2.66 \times 10^{11}$~pairs/mW is obtained using Equation~\ref{eq:multimodespectraldensity}. However, using Equation~\ref{eq:spectraldensity}, the theoretical efficiency using a single frequency pump is calculated to be $1.28 \times 10^{12}$~pairs/mW. Therefore, approximately an order of magnitude of enhancement is expected by using a single frequency pump over a multimode pump.

\bibliography{supplementary}